\documentclass[utf8]{frontiersFPHY} 

\setcitestyle{square} 
\usepackage{url,hyperref,microtype,subcaption}
\usepackage[onehalfspacing]{setspace}

\newcommand{\lb}{\lambda_\text{deB}}

\newcommand{\mb}{\mathbf}
\newcommand{\p}{\partial}
\newcommand{\tx}{\textit}

\newcommand{\beq}{\begin{equation}}
\newcommand{\eeq}{\end{equation}}
\newcommand{\bdi}{\begin{displaymath}}
\newcommand{\edi}{\end{displaymath}}
\newcommand{\beqn}{\begin{eqnarray}}
\newcommand{\eeqn}{\end{eqnarray}}
\newcommand{\f}{\frac} 

\newcommand{\MS}{M_{\odot}}

\newcommand{\Rv}{R_{\rm vir}}
\newcommand{\Mv}{M_{\rm vir}}
\newcommand{\vv}{v_{\rm vir}}
\newcommand{\Tlb}{T_\text{deB}}

\def\keyFont{\fontsize{8}{11}\helveticabold }
\def\firstAuthorLast{T. Rindler-Daller} 
\def\Authors{Tanja Rindler-Daller\,$^{1,*}$} 
\def\Address{$^{1}$Institut f\"ur Astrophysik, Universit\"atssternwarte Wien, Universit\"at Wien, Vienna, Austria }
\def\corrAuthor{Corresponding Author}
\def\corrEmail{tanja.rindler-daller@univie.ac.at}

\begin{document}
\onecolumn
\firstpage{1}

\title[Quantum-coherent DM in the Milky Way]{To observe, or not to observe, quantum-coherent dark matter in the Milky Way, that is a question} 

\author[\firstAuthorLast ]{\Authors} 
\address{} 
\correspondance{} 

\extraAuth{}

\maketitle

\begin{abstract}

\section{}
In recent years, Bose-Einstein-condensed dark matter (BEC-DM) has become a popular alternative to standard, collisionless cold dark matter (CDM). This BEC-DM -also called scalar field dark matter (SFDM)- can suppress structure formation and thereby resolve the small-scale crisis of CDM for a range of boson masses. However, these same boson masses also entail implications for BEC-DM substructure within galaxies, especially within our own Milky Way. Observational signature effects of BEC-DM substructure depend upon its unique quantum-mechanical features and have the potential to reveal its presence. Ongoing efforts to determine the dark matter substructure in our Milky Way will continue and expand considerably over the next years. In this contribution, we will discuss some of the existing constraints and potentially new ones with respect to the impact of BEC-DM onto baryonic tracers. Studying dark matter substructure in our Milky Way will soon resolve the question, whether dark matter behaves classical or quantum on scales of $\lesssim 1$ kpc.

\tiny
 \keyFont{ \section{Keywords:} cosmology, Bose-Einstein-condensed dark matter, galactic halos, Milky Way, dark matter substructure, quantum measurement} 
\end{abstract}

\section{Introduction}\label{sec:intro}

According to the theme of this Research Topic article collection, we might imagine a fictitious conversation in the Parnassos of deceased scholars, involving Rubin, Einstein and Planck:
while Rubin would elaborate on her observational findings of dark matter (DM) from the dynamics of galaxies, the question will arise whether we understand gravity sufficiently well to explain this phenomenology. Einstein would calmy point out that general relativity has passed all tests so far, including ever newer ones from the detection of gravitational waves from black hole mergers, to the direct observation of the central supermassive black hole of galaxy M87, and to tests of the equivalence principle, and so forth. As Einstein has convinced (almost) everyone, Planck may finally make a case that quantum mechanics may also have a say, after all, in this discussion. 
This is the framework of this paper: we believe in particle dark matter, consider general relativity the correct theory of gravity (at low energies), but quantum mechanics will also play a role. The wave-particle dualism applies also to DM particles, and it is appropriate to treat many DM candidates in the particle regime. In this paper, we consider DM whose wave nature cannot be neglected on astronomical or galactic scales anymore, i.e. quantum properties can affect astrophysical DM phenomenology.
Ultralight bosonic DM with particle masses of $m \sim (10^{-23} - 10^{-18})$ eV/$c^2$ belongs to this category, and the literature has seen a recent explosion in the interest for this candidate, under the header of ``scalar field dark matter (SFDM)'', ``fuzzy dark matter (FDM)'', ``wave dark matter'', ``ultralight axion DM'', ``Bose-Einstein-condensed dark matter (BEC-DM)'', etc.
Early works on this subject include \cite{Sin94,Peebles2000,Goodman2000,sf4,sc19,2000PhRvL..85.1158H}, though microscopic details of the considered models may differ. Reviews on SFDM can be found e.g. in \cite{rev1,UL2019}, on axions e.g. in \cite{rev3}.     

We will focus in this paper on the simplest model of a single bosonic species in a BEC, which has \textit{no self-interactions} and no direct coupling to standard model particles. Also, this paper will focus on BEC-DM substructure within the Milky Way. 

The mathematical framework described below applies to any DM bosons which can form a BEC, whether the underlying bosons be ultralight, appropriate for FDM, or whether they be QCD axions with $m \sim (10^{-6}-10^{-5})$ eV$/c^2$. Yet, we stress that BEC-DM need not be an axion-like particle! In fact, as we will see, the formalism of equ.(\ref{gp}-\ref{norm}) below works right away, if we start with the premise that the scalar field be complex. On the other hand, axions are real pseudo-scalars, so they need to be transformed to a complex field, a procedure which is valid only if self-interactions are neglected, see \cite{BMZ18}. However, since all constraints that we will mention build upon that formalism, we will consider them as equally applicable, not discussing any potential differences brought about by varied microscopic BEC-DM implementations (see also footnote 2).


The lower the boson mass, the larger the expected substructure in BEC-DM and its impact onto small-scale structure formation. Consequently, cosmological probes have been used to infer lower bounds on that mass. Those bounds are weakened, if BEC-DM constitutes only part of the DM, but we are interested here in constraints that assume that all of the DM is in the form of BEC-DM. Previous constraints covered a broad range of lower-bound estimates. CMB anisotropies require $m > 10^{-24}$ eV/$c^2$ \cite{2015PhRvD..91j3512H}. The distribution of globular clusters in Fornax demands $m \sim (0.3-1)\times 10^{-22}$ eV/$c^2$ \cite{free_const3}, while the half-light mass of ultrafaint dwarf galaxies requires $m \sim (3.7-5.6)\times 10^{-22}$ eV/$c^2$ \cite{sc20}.
More recent papers tend to come up with tighter bounds.
For example, the ultrafaint galaxy Eridanus II features in the bounds by \cite{MN19}, where the requirement for it to be a subhalo today mandates $m \gtrsim 8\times 10^{-22}$ eV$/c^2$. If the age and size of its prominent star cluster is also taken into account, the range $8\times 10^{-22}~ \text{eV}/c^2 \lesssim m \lesssim 10^{-19}~\text{eV}/c^2$ is disfavoured, in addition. Similar bounds of $m \gtrsim 2.9\times 10^{-21}$ eV/$c^2$ follow from the study by \cite{Nadler21}, which also imposes the requirement of matching subhalo mass functions to the Milky Way satellite population, including ultrafaint dwarfs. Moreover, \cite{Safarzadeh_2020} consider the competing demands from different dwarf galaxy populations, excluding masses smaller than $6 \times 10^{-22}$ eV/$c^2$. 
These recent studies seem to converge to values inferred earlier from 
the Lyman-$\alpha$ forest, which sticked out for some time concerning its limits of $m \gtrsim (20-30)\times 10^{-22}$ eV/$c^2$ \cite{sc22,sc23}.
Further, recent constraints have been summarized in \cite{AWP} for ultralight axions; most of these apply to bosons more generally, as well.



Now, this paper focuses on the study of DM substructure within the Milky Way, a topic which is being revolutionized by modern astrometry missions like GAIA, and even the near future holds promise in our quest to reveal the nature of DM. Unlike the measurement of the rotation curves of external galaxies, this time our inside-perspective of the Milky Way will come to our advantage. The high-precision big data from astrometry missions will help in multiple ways with respect to the DM problem, e.g. i) determine accurately the total mass of the Milky Way, ii) detect DM substructure indirectly from dynamical inferences and constrain it in the process, iii) reveal and disentangle the complicated astrophysics of baryonic tracers in the Milky Way so that we better understand the impact of different baryonic components and their interaction with DM, etc.

That the properties of many tracers such as open clusters in the disk, or the interstellar medium, have been studied and ``understood'' in all these decades without the need to account for DM substructure could be regarded somewhat as a mystery. The fact that it \textit{appears} that many researchers of the Milky Way do not need to worry in their work about DM substructure should alert cosmologists. What shall that mean for the nature of DM? Standard CDM predicts a plentitude of small-scale structures and subhalos, e.g. for WIMPs (weakly interacting massive particles) all the way down to Earth-mass scales, and alternative DM models also predict enough substructure for it to ``matter''.
Sure enough, the quest to study DM substructure and its potential impact onto the disk, or tracers within it, dates back at least to the early days of CDM (some references will be given below). 
But somehow it seems that the overwhelming number of individual self-gravitating DM subhalos (we call them ``drops'' in this paper for reasons discussed below) should be either so tiny that their individual dynamical fingerprints are negligible, or else so large that a given single drop covers big stretches of the Milky Way disk or halo, in such a way that the encompassed stellar tracers hardly feel its presence over most of their lifetimes. Too many ``intermediate size'' DM drops might be prone to cause too much dynamical chaos in the Milky Way and to its stellar tracers\footnote{Not to speak of the greater impact onto the spiral structure, or the central galactic bar, which can be also used to constrain DM; we will not discuss these here. }, such as giant molecular clouds (GMCs), large-scale gas filaments, open clusters or binaries. Thus, there is still work to be done to establish how much and what DM substructure is in accordance with the observed properties of the baryonic components of the Milky Way.

This paper is organized as follows. The equations of motion of BEC-DM, some basic concepts and halo structure will be discussed in Section 2. Novel aspects of BEC-DM, especially with respect to its substructure and the impact of quantum phenomena onto stellar tracers in the Milky Way will be reviewed in Section 3.  In Section 4, we will discuss further quantum aspects of BEC-DM, which add even more distinction to the standard CDM model.


\section{Dynamics, characteristic scales and halos at large}

We assume that BEC-DM consists of fundamental spin-$0$ bosons, all with the same mass $m$. There are ways to create these bosons in the early Universe, either thermally, or via a vacuum realignment mechanism akin to the QCD axion. We will not consider these early times, but only start with the premise that these bosons are ``cold'' (i.e. nonrelativistic) by the time cosmic structure forms and ever since. In addition, we focus on scenarios where these bosons are (almost) all in the same quantum state, and they got there by a Bose-Einstein phase transition\footnote{There is still confusion in the literature, which does not distinguish a squeezed state from the ground state of a Bose-Einstein-condensed system, but we will not enter this discussion here and presume that condensation actually happened at some point in the early Universe.}.
Once we have (almost) all $N$ bosons in that same quantum state, we can describe their evolution through a single scalar field $\psi$ (hence SFDM), i.e. the dynamics of the original $N$-particle wavefunction reduces to a $1$-particle "wavefunction" $\psi$ of the Bose-Einstein condensate (BEC) itself.
Its equation of motion is the Gross-Pitaevskii (GP) equation, a form of nonlinear Schr\"odinger equation, which is coupled to the Poisson equation in order to describe self-gravitating systems. The combined system of equations\footnote{In models without self-interaction, as described here, these equations of motion are also known as Schr\"odinger-Poisson (SP) system.} is abbreviated GPP,
\begin{equation} \label{gp}
 i\hbar \frac{\partial \psi}{\partial t} = -\frac{\hbar^2}{2m}\Delta \psi + m\Phi \psi,
 \eeq
 \beq \label{poisson}
  \Delta \Phi = 4\pi G m |\psi|^2,
   \end{equation}
    where
$|\psi|^2(\mathbf{r},t) = n(\mathbf{r},t)$ is the number density of dark
matter bosons of mass $m$ and $\Phi(\mathbf{r},t)$ is the gravitational
potential of the object. 
If we consider a single self-gravitating BEC-DM object, say an isolated galactic halo or a substructure DM drop within a bigger host halo, we can normalize the density to account for the $N$ bosons of that particular object, i.e.
\beq \label{norm}
    \int_V |\psi|^2 = N.
     \eeq
The above equations are appropriate for isolated objects, but at cosmological scales one would need to add terms due to Hubble expansion. 
This model as decribed here is sometimes called the simplest "vanilla" model of ultralight bosonic BEC-DM: it is modelled using a plain scalar field.  
Importantly, the vanilla types have no internal boson self-interactions, and in this context they have been
called fuzzy DM, $\psi$DM, wave DM or free SFDM. In this work, we will call the model often fuzzy dark matter (FDM), in observance of the early paper
by \cite{2000PhRvL..85.1158H}, wherein that term was coined, although fiducial FDM models usually require ultralight bosons, $m \sim (10^{-23}-10^{-22})$ eV$/c^2$, in order for quantum phenomena to be visible on galactic scales (i.e. the mere absence of self-interaction does not constitute FDM as it was originally conceived). 
We stress that SFDM encompasses much more varieties\footnote{For instance, adding a 2-boson self-interaction to the equations of motion expands the parameter space of models, because in addition to $m$ we gain another parameter to describe this self-interaction. For example, an effective contact interaction in the limit of vanishing energy can be modelled with a constant coupling strength $g$, yielding a term $g|\psi|^2\psi$ on the right-hand side of (\ref{gp}), which corresponds to a quartic term $\propto g|\psi|^4$ in the potential energy. The sign of $g$ determines whether self-interaction is attractive or repulsive. On the other hand, axion-like particles have a cosine interaction potential which, upon expansion, will also result in such a quartic term. The axion self-interaction is usually attractive, but mostly neglected in the literature, altogether.  Again, we disregard models with self-interaction in this paper, as well. } than just the simplest cases considered here.


The GPP formalism assumes that all particles of the object under consideration are in the BEC and the latter requires that occupation numbers are high, and $n\lb^3 \gg 1$ be fulfilled, where $n$ is again the number density of bosons. $\lb$ is the wavelength of the boson assigned by Louis de Broglie in his wave-particle dualism, the famous de Broglie wavelength, 
\beq \label{deB}
\lb = \f{h}{p} = \f{h}{mv},
\eeq
assuming that the boson moves with nonrelativistic velocity $v$. It may not be so obvious what $v$ should be in our context\footnote{In de Broglie's wave-particle dualism, $v$ is understood as the group velocity of the stationary wave ``assigned'' to \textit{a} particle, i.e. the velocity of \textit{this} particle is the group velocity of that wave.}. In a laboratory setting, where we think of cooling a thermal atomic gas, say, to become a BEC, we have the thermodynamic temperature $T$ which refers to the average kinetic energy in a thermalized sample of atoms, $E_\text{kin}=mv^2/2 \approx k_B T$. Solving for $v$ and inserting in (\ref{deB}), we have the \textit{thermal} de Broglie wavelength $\lb = h/\sqrt{2m k_B T} \propto T^{-1/2}$. Now, in the high-$T$ regime, the atoms move like ``billard balls''; at low-$T$ the individual atomic wave packets (matter waves) of size $\lb$ become increasingly larger. At the critical temperature of condensation, $T=T_c$, $\lb$ becomes of the order of the interatomic separation, atomic wave packets overlap and individual particles become indistinguishable, and for $T \to 0$, we end up with a giant matter wave, a pure BEC. Therefore, if we apply the GPP formalism to DM bosons by using these equations all the way back to the onset of structure formation, we implicitly presume that condensation had to happen in the earlier Universe, and not during halo formation, i.e. halos do not form via condensation to a BEC. Instead, overdensities form within a homogeneous BEC-DM background and halos arise out of these overdensities in the process of nonlinear gravitational collapse. Nevertheless, $\lb$ in halos of BEC-DM today should respect the nonrelativistic regime of those halos, i.e. if we say that these halos are held up in equilibrium against gravity by a ``dispersion pressure'' of non-relativistically moving particles (akin to stars or CDM particles in galaxies), we could argue that the virial temperature $T_\text{vir}$, i.e. the mean temperature at which a gravitationally bound system will satisfy the virial theorem, of the bosons takes the role of $T$ above, respective the virial\footnote{Just like for CDM, the ``dynamic temperature'' of BEC-DM increases, once originally uncollapsed matter finds its way into post-collapse halos.} velocity $\vv \approx \sqrt{G\Mv/\Rv}$ should be used in the definition of $\lb$, resulting in 
\beq
\lb^2 = \left(\f{h}{m}\right)^2 \f{\Rv}{G\Mv}.
\eeq
For BEC-DM models for which $\lb$ can be as large\footnote{It would be more meaningful to assign $\lb$ to the diameter of the object, but since much of the literature assigns the radius, instead, we also stick to this choice.} as the virial radius of a self-gravitating object, $\lb \approx \Rv$ (which is the case, if boson self-interaction is neglected, in particular for FDM), we have
\beq \label{mass-radius}
\Rv = \left(\f{h}{m}\right)^2 \f{1}{G\Mv} \approx 40 \left(\f{\hbar}{m}\right)^2 \f{1}{G\Mv}.
\eeq
In fact, early works such as \cite{Membrado} or \cite{GU04}  calculated already the numerical profile of an isolated, gravitationally bound ground-state system (``soliton''), as a result of solving the above differential equations as an eigenvalue problem. The corresponding mass-radius relationship reads as 
\beq \label{massradius}
R_{99} = 9.946 \left(\f{\hbar}{m}\right)^2 \f{1}{GM},
\eeq
where $R_{99}$ includes $99\%$ of the mass $M$ of the object; the system has no compact support, hence the radial cutoff, see also\footnote{Analytical approximations to the numerically calculated soliton profile, like a Gaussian profile or an $(n=2)$-polytrope as considered in \cite{2019PhRvD.100l3506C}, yield the same relationship (\ref{massradius}), but with a different factor in front of order $5$, see e.g. equations 64 and 72 in \cite{SRS21}.} \cite{Chavanis2011}. This confirms \tx{a posteriori} that it was reasonable to evaluate $\lb$ using a typical (circular, virial, dispersion) velocity of the ``halo'', if $\lb$ extends over much of the system in question, that gave us (\ref{mass-radius}) which is ``close'' to (\ref{massradius}).
These solitons are attractor solutions of the GPP (or SP) system of equations: in the process of long-time virialization, scalar-field mass is expelled to infinity (a process termed ``gravitational cooling'' in \cite{SS94}), leaving behind an isolated soliton. Of course, continuous mass infall and mergers come in the way and gravitational cooling can be delayed substantially, especially for FDM halos at large, as described below.

Notice that (\ref{massradius}) is valid for any boson mass $m$, ultralight or not, but the size of the soliton shrinks quickly for rising boson mass.  Since we also discuss boson masses intermediate between ultralight and QCD axion-like, we will call self-gravitating objects, which follow the relation (\ref{massradius}), ``solitonic drops'', or ``drops'' for short.

However, these ground-state solitons not only describe individual BEC-DM drops, they also represent the central parts of BEC-DM halos, which formed from mergers either in a static or expanding background, the so-called ``solitonic cores'', whose size is of order (\ref{massradius}). This has been first convincingly shown by the FDM simulations of \cite{SCB14, Schive14}, and the small boson mass of fiducial FDM models, $m \sim 10^{-22}$ eV$/c^2$, makes these cores visible to simulations, given their resolution limits. 
Now, the relation in (\ref{massradius}), valid for solitons, fails to describe galactic halos in general, i.e. halos extending beyond the solitonic core. This shortcoming has been realized for a while, see e.g. \cite{RS14}. Also, (\ref{massradius}) implies that the density of solitons increases sharply, the more mass they assemble, for their size shrinks in inverse proportion. Detailed comparison to observations have just been finding more ``symptoms'' of this property of the soliton, applied to solitonic halo cores, questioning whether BEC-DM in the form of FDM is really a cure to the small-scale crisis, see \cite{VBB19}. \\
More simulations of FDM by \cite{SNE16} or \cite{Mocz17}, following \cite{SCB14}, have confirmed that BEC-DM halo formation leads to the appearance of an ``envelope'' which can extend in size very much beyond $\lb$, i.e. even if the size of the halo core is of order $\lb$, the halo at large, i.e. its envelope can be much bigger, such that $\Rv \gg \lb$, after all\footnote{Indeed, even in gravity-free laboratory settings, it has been known that the BEC state remains robust, in spite of splitting the condensate, or upon expansion of the condensate that increases even the interparticle separation beyond $\lb$.}.
Explanations of this large-scale dynamics of FDM can be found e.g. in \cite{Mocz18}, or \cite{Taha1}. Simulations find that $\lb$ should be evaluated using a typical velocity of the core. If the latter is close to that of the halo at large, i.e. if $v_{core} \approx \vv$ (``velocity dispersion tracing'') as has been pointed out by \cite{2019PhRvD.100h3022C}, or \cite{Bar18}, then this ``tracing'' implies certain relationships between the mass and radius of the halo and its core, see \cite{2019PhRvD.100l3506C, Padilla21}. 

The smaller the boson mass, the more prominent the solitonic cores at the centers of halos, with observational consequences. Their impact on galactic rotation curves has been studied in \cite{Bar18} with the conclusion that masses of $m \sim (10^{-22}-10^{-21})$ eV/$c^2$ are in tension with data of nearby galaxies. The impact of a central soliton onto the ``central molecular zone'' in the Milky Way has been specifically addressed by \cite{Li_2020}, finding that a core of mass $\approx 4\times 10^8 \MS$ and radius $\approx 0.05$ kpc could fit the data. Such a core requires $m \sim (2-7)\times 10^{-22}$ eV/$c^2$.

Apart from the core-envelope structure of halos, some FDM simulations find large-scale coherence effects around and between halos (\cite{SCB14}, \cite{Mocz17}), extending all the way to cosmic filaments \cite{Mocz20}, as well as finding signs of quantum turbulence within halo envelopes \cite{Mocz17}.
Unfortunately, current simulations cannot resolve the DM substructure expected in FDM and BEC-DM, in general. Nevertheless, we can infer some implications described in the next section.

Before that, we summarize a few more basic properties of BEC-DM. First, we state the fundamental scales of the bosons under question in fiducial units for an easier comparison later.  We have $\lb$ as already discussed above, which is
\beq \label{char1}
\lb = \f{h}{mv} = 1.2 ~\text{kpc} \left(\f{10^{-21} ~\text{eV}/c^2}{m}\right)\left(\f{10 ~\text{km}/s}{v}\right).
\eeq
The time for a boson to cross 
the length scale $\lb$ is called ``coherence time'',
\beq \label{char2}
\Tlb = \f{h}{mv^2} = 118,124 ~\text{yr} \left(\f{10^{-21} ~\text{eV}/c^2}{m}\right)\left(\f{10 ~\text{km}/s}{v}\right)^2.
\eeq
This time scale is related to the coherence properties of BEC-DM as a giant matter wave.
The free-fall time of an object with mean density $\bar{\rho}$, defined as $T_\text{ff} = 1/\sqrt{\pi G \bar{\rho}}$, is the same, $T_\text{ff} = \Tlb$, if\footnote{See \cite{RS12}, where $\lb = \Rv$ amounts to $\Omega_\text{grav} = \Omega_{QM}$, where $\Omega_\text{grav} = 1/T_\text{ff}$, and 
$\Omega_{QM} = \hbar/(m\Rv^2)$ is the angular frequency of a uniformly rotating object with mass $M$ and angular momentum $L_{QM} = N\hbar = M\Rv^2\Omega_{QM}$. Thus, $1/\Omega_{QM}$ corresponds to $\Tlb$ up to a factor of $2\pi$.} $\lb = \Rv$. However, in case of a halo core for which $\lb \ll \Rv$, its dynamical time scale $T_\text{ff}$ is much longer than its coherence time $T_\text{deB}$, i.e. intuitively speaking, any boson could cross the core many times before any dynamical effect onto the core, imparted by the halo at large, is being ``felt''by that boson.  

Now in addition, by de Broglie's wave-particle dualism, a frequency $\nu$ (an ``internal clock''), is assigned to any particle with mass $m$ by setting $mc^2 = h\nu$ in an attempt to merry relativity with quantum mechanics\footnote{De Broglie was aware of the fact that this marriage was not relativistically invariant, prompting his actual conception of the particle's stationary wave. Since $v \ll c$ for halo bosons, we disregard the distinction between frequencies in the rest or moving frame.}, i.e. if only rest-mass matters in determining this frequency, the usual dispersion relation of a free particle results, $\nu = m c^2/h$, or equivalently $\omega = m c^2/\hbar$.
From this well-known procedure follows immediately the Compton wavelength, which is
\beq \label{compton}
\lambda_0 = \f{c}{\nu} = \f{h}{mc} = 0.04 ~\text{pc} \left(\f{10^{-21} ~\text{eV}/c^2}{m}\right).
\eeq
The ``clock time'' associated with the oscillation frequency is thus
\beq \label{char3}
T_0 = \f{1}{\nu} = \f{h}{mc^2} = 0.132 ~\text{yr} \left(\f{10^{-21} ~\text{eV}/c^2}{m}\right).
\eeq
The quantities\footnote{Sometimes these quantities are defined in terms of $\hbar$, which amounts to differences of a factor of $2\pi$ everywhere.} in (\ref{char1}-\ref{char3}) are the characteristic time and length scales for condensed DM bosons without other microphysical properties, especially for those without self-interaction.
The latter scales are much smaller than the former, $\lambda_0 \ll \lb$, $T_0 \ll \Tlb$, i.e. for physical phenomena on scales above $\lb, \Tlb$, one can safely average over these ``Compton scales''. However, $\lambda_0$ and $T_0$ can still be of macroscopic size, albeit unimportant on structure-formation scales, with consequences discussed in Section 4. In fact, to put into perspective the above fiducial numbers for $m = 10^{-21}$ eV$/c^2$, the corresponding numbers for the QCD axion, choosing $m = 10^{-5}$ eV$/c^2$ and $v=200$ km/s at the solar circle, yield $\lb = 185$ m, $\lambda_0 = 0.12$ m, $T_\text{deB} \approx 10^{-6}$ sec, and $T_0 \approx 4\times 10^{-10}$ sec, respectively.

Notice that the ``Compton scales'' are natural constants for any given BEC-DM model, whereas the ``de Broglie scales'' depend upon the environment of bosons, i.e. for the latter it matters whether bosons belong to the uncollapsed fraction, or be part of a halo overdensity or substructure, respectively. This will be important to remember.

We stress that it is possible to rewrite the GP equation in a form that recovers hydrodynamical conservation equations of continuity and momentum (``quantum hydrodynamics''), which is just a different way of saying that we can use the de Broglie-Bohm formulation of quantum mechanics.
Introducing a polar decomposition (also known as Madelung transformation in this context), according to 
\beq \label{polar}
\psi(\mb{r},t) = |\psi|(\mb{r},t)e^{iS(\mb{r},t)} =
\sqrt{\f{\rho(\mb{r},t)}{m}}e^{iS(\mb{r},t)},
 \eeq
 with the BEC-DM mass density $\rho = m|\psi|^2$,
and observing that the quantum-mechanical current density,
\bdi
 \mb{j}(\mb{r},t) = \f{\hbar}{2im}(\psi^* \nabla \psi - \psi \nabla
 \psi^*) = n(\mb{r},t)\f{\hbar}{m}\nabla S(\mb{r},t)
  \edi
can be expressed in terms of the (bulk) velocity, if we write
    $\mb{v} = \hbar \nabla S/m$,
    then equation (\ref{gp}) can be written in the form
\beq \label{fluid}
     \rho \f{\p \mb{v}}{\p t} + \rho (\mb{v} \cdot \nabla)\mb{v} =
      -\rho \nabla Q
    - \rho \nabla \Phi,~~~\mbox{where}~~~ Q = -\f{\hbar^2}{2m^2}\f{\nabla^2 \sqrt{\rho}}{\sqrt{\rho}}
     \eeq
     and
 \beq \label{hd3}
\f{\p \rho}{\p t} + \nabla \cdot (\rho \mb{v}) = 0.
 \eeq
We can see that (\ref{hd3}) has the form of a continuity equation, while (\ref{fluid}) has the form of an Euler-like fluid equation of motion. Hence, the original GP equation has been re-written as a set of quantum hydrodynamical equations\footnote{The equivalence of these pictures has been questioned in \cite{W94}, but that claim has been shown to be invalid in \cite{H10}.} with ``hydro variables'' $\rho$ and $\mb{v}$.
The quantum potential $Q$ gives rise to a so-called ``quantum pressure'' on the right-hand-side of the fluid momentum equation in (\ref{fluid}). It stems basically from the quantum-mechanical uncertainty principle. 
The quantum hydrodynamical equations are then solved self-consistenly, along with the Poisson equation (\ref{poisson}).

The astrophysics literature uses both formalisms, and the choice for one over the other is influenced by taste and numerical method at hand. 

\section{Aspects of BEC-DM substructure}

\subsection{Wave interference}

The question of DM substructure is a critical one, for CDM and for alternative DM models. In this paper, we discuss certain novel aspects of BEC-DM substructure, which have been increasingly studied in the literature. A review can be found in \cite{Niemeyer20}, especially focusing on the cosmological perspective and axion-like particles. 
Estimating the minimum size of gravitationally bound objects in a given DM model can be done straightforwardly, in principle, by applying a Jeans analysis. Yet, the dynamical details depend upon the actual particle physics models and are complicated to work out case by case. The abundance of expected substructure is even harder to estimate because that depends upon the primordial structure formation on small scales, followed by the highly nonlinear dynamics of merging, gravitational heating, tidal stripping, etc. Many of these issues require future studies.  

For BEC-DM, we encounter additional phenomena, relevant to the question of substructure, due to its quantum nature.
Quantum wave interference phenomena arise on galactic scales, which have been studied for FDM using simulations and some analytic estimates.
We will discuss some of these findings, and otherwise restrain to crude estimates to guide our discussion of substructure in the Milky Way. 

It has been known for a long time that DM substructure can potentially impact the amount of dynamical heating of the Milky Way thick disk, see e.g. \cite{LO85}. 
There is a beautiful study by \cite{Church19}, which is concerned with the impact of FDM substructure on the thickening of the Milky Way disk. The central part of their Galactic halo model comprises a solitonic core, outside of which the averaged density profile would follow a CDM-like profile, more precisely an NFW profile \cite{NFW}.
In the process, the authors discuss the qualitative differences between 
virialized FDM subhalos (``drops'', thus the collapsed fraction in substructure) versus non-virialized, but overdense FDM ``wavelets'' (uncollapsed fraction in substructure). The latter arise as a result of linear interference patterns of the underlying velocity dispersion of bosons, which has been seen in simulations of idealized halo formation and virialization, e.g. in \cite{Mocz17}, and can be associated with the quantum phenomena inherent to BEC-DM/FDM. As such, the abundance of these wavelets increases sharply with decreasing boson mass, as expected; by the same token wavelets are strongly suppressed by a factor $\left(\f{m}{10^{-22} ~\text{eV}/c^2}\right)^{-3}$ for increasing boson mass, as BEC-DM in general resembles CDM, the higher $m$. In general, FDM drops are more easily disrupted in the inner parts of the host halo, therefore wavelet dynamics dominates the central parts, while the dynamics of FDM drops is relevant in the outer parts of the halo. As such, the heating of the disk will be dominated by wavelets in the inner parts - which includes the solar circle at $r \approx 8$ kpc-, for models with low enough boson mass.  
Leaving aside details, \cite{Church19} estimate a lower bound on $m$ by requiring that FDM wavelets should dynamically heat the oldest disk stars in the solar neighborhood to the observed value of $\sim 30$ km/sec within a Hubble time.  They find $m > 6\times 10^{-23}$ eV$/c^2$.
Their model also reproduces an age-velocity dispersion relation, $\sigma \propto t^{0.4}$, close to the observed one. Of course, given that other galactic components, e.g. spiral arms, are also expected to contribute to disk heating, that lower bound on $m$ is pessimistic.      

Apart from disk heating, FDM wavelets have been also constrained by their impact onto the thickening of thin stellar streams by \cite{AL18}. Since stellar streams are even ``colder'' than the disk, the resulting bound on the boson mass is stricter, namely $m > 1.5\times 10^{-22}$ eV$/c^2$.

Somewhat earlier dynamical studies of FDM substructure, including derivations of relaxation times and investigations of the impacts onto multiple baryonic tracers, can be found in the works of \cite{Hui17} and \cite{BarOr19}. (Having at hand a copy of \cite{BT2008} helps in studying the papers). \cite{Hui17} derive a relaxation time for what they call ``FDM quasiparticles'' of mass 
\beq \label{Huimass}
m_\text{eff} \sim \rho\left(\f{\lb}{2}\right)^3 \sim 3\times 10^5 ~\MS \left(\f{10^{-22} ~\text{eV}/c^2}{m}\right)^3,
\eeq
using values appropriate for the solar circle, $r \sim 10$ kpc, $v \sim 200$ km/s, and $\rho \sim 0.01~ \MS ~\text{pc}^{-3}$ for the local DM density\footnote{See e.g. \cite{Bovy12} who estimate $\rho_{DM} = 0.008 \pm 0.003 ~\MS/\text{pc}^3 = 0.3 \pm 0.1$ GeV/cm$^3$.}. 
These quasiparticles can be thought of as ``linear overdensities'' within the galactic background, i.e. they are not solitonic drops, but similar to the wavelets discussed before. The relaxation time of such quasiparticles, or any DM substructure, depends upon the distance from the galactic center, and the general expression of \cite{Hui17} reads as
\beq \label{Huirelax}
t_\text{relax}(r) \sim \f{0.4}{f_\text{relax}}\f{m^3 v^2 r^4}{\pi^3 \hbar^3} \sim \f{10^{10} ~\text{yr}}{f_\text{relax}} \left(\f{r}{5 ~\rm{kpc}}\right)^4 \left(\f{10^{-22} ~\text{eV}/c^2}{m}\right)^{-3}\left(\f{v}{100 ~\rm{km}/s}\right)^2
\eeq
with a dimensionless constant $f_\text{relax} \lesssim 1$.
This estimate assumes $t_\text{\relax} \approx 0.1 ~t_\text{cross}~ M/m_\text{eff}$, with the crossing time $t_\text{cross} = r/v$. We can see that the relaxation time is about a Hubble time for fiducial FDM boson masses at sufficiently large $r$. 

\cite{BarOr19} conduct an accurate study of relaxation of FDM substructure and dynamical heating, with an emphasis on the question how long it takes to bring infalling massive objects to acquire the velocity dispersion of the host halo. They adopt a singular isothermal sphere for the Galactic halo\footnote{\cite{Bovy13} estimate the profile for the DM density at the solar circle $R_0$ according to $\rho_{DM}(r\approx R_0) \propto 1/r^{\alpha}$, where $\alpha < 1.53$ at $2\sigma$.}, which explains some of the deviations between their's and the results of others'. To a certain extent, they confirm the previous estimates by \cite{Hui17}, e.g. the mass of 
the quasiparticle/wavelet, if evaluated using the mean density of the quasiparticle,
\beq \label{BarOrmass}
m_\text{eff} = \sqrt{\f{\pi}{2}}\f{\hbar^3}{Gm^3 v r^2} = 1.03\times 10^7 ~\MS
\left(\f{r}{\rm{kpc}}\right)^{-2} \left(\f{10^{-22} ~\text{eV}/c^2}{m}\right)^{3}\left(\f{v}{200 ~\rm{km}/s}\right)^{-1}.
\eeq
Detailed simulations including baryons will be required to see how the complicated wave interference in BEC-DM/FDM halos can be in accordance with observational findings regarding the dynamical heating of the disk. In this regard, widely extended stellar streams far from the disk will be a ``cleaner'' probe in order to study the impact of wave interference or solitonic drops onto them. The previously mentioned studies already have shown that DM in the Milky Way cannot be ``too quantum'' on scales $\gtrsim 1$ kpc, i.e. the boson mass cannot be too low, for otherwise we would have inferred its presence already. We stress that these studies with their conclusions provide independent constraints from those which refer to the small-scale problems at the scales of dwarf galaxies and their bounds on FDM boson mass, as described in Section 1.

Another increasingly studied phenomenon is the nature and impact of dynamical friction of BEC-DM and how it differs from the classic approaches used for CDM. Previous estimates and considerations can be found in \cite{Hui17} applied to FDM, but the question has been taken up in \cite{Lancaster20}, also for FDM, in much more analytic and numerical detail, including applications to various systems such as the globular clusters in the Fornax dwarf spheroidal, or the Sagittarius stream. The authors investigate dynamical friction acting on a satellite which moves through an FDM background.
They identify basically three distinct regimes in FDM dynamical friction: i) $\lb$ is large and the wake created in the process is set by the quantum pressure, then dynamical friction is well described by linear perturbation theory; ii) the background has a velocity dispersion, $\lb$ is small, then the wake behaves similarly to a classical Chandrasekhar wake; and iii) the length scales of the wake and $\lb$ of the velocity dispersion are comparable, then the wake has a stochastic character, arising from interference crests of the background, with overdensities and underdensities strongly influencing the motion of the satellite, and the dynamical friction force becomes uncertain. This regime requires detailed simulations, but it offers potential new observational signature effects, given that the wave-like transient nature of the wake is yet another quantum-coherent impact of BEC-DM onto baryons.      
Again, the prominence of these novel impacts depends upon the boson mass, and the higher the latter, the less we will be able to recognize these quantum phenomena. In addition, \cite{Lancaster20} find that FDM with $m \gtrsim 10^{-21}$ eV$/c^2$ would not resolve the so-called timing problem of the globular clusters of the Fornax dwarf galaxy, i.e. the fact that we would have expected these clusters to have fallen into Fornax within a Hubble time given CDM, in accordance with earlier work \cite{free_const3}. In order to resolve this dynamical conundrum, fiducial FDM models with small enough boson mass would be required, and these are hit hard by other constraints, as mentioned before.

\subsection{Solitons abounding}

Here, we turn to solitonic drops in anticipation that BEC-DM bosons may have to have higher masses than originally hoped. We have already mentioned that the importance of subhalos, i.e. drops, increases with mass $m$, while the importance of wavelets decreases. Therefore, we will put into perspective their potential impact onto baryonic tracers.

The size of gravitationally bound BEC-DM drops is determined by $\lb$, because structure formation below $\lb$ will be suppressed and structure above $\lb$ is not of minimum size anymore, though this statement is crude and requires clarification. Suppose we want to have a minimum size of isolated galactic halos in gravitational equilibrium, supposedly those which host the smallest galaxies. This requirement implies a lower bound on the boson mass $m$, as follows. For the sake of discussion, we choose a fiducial halo\footnote{The results by \cite{MN19} and \cite{Nadler21} suggest that our fiducial choice is already ruled out, but detailed contraints of factors of a few are not so decisive for the general arguments made here.} appropriate to host an ultrafaint dwarf galaxy with virial velocity of $10$ km/s and virial radius of $\lb^{mingal} = 1.2$ kpc, which requires a boson mass of $m=10^{-21}$ eV/$c^2$, see (\ref{char1}). The superscript simply refers to the value for the de Broglie length from the requirement of producing that minimum galactic halo size \textit{today}. 
Lower $m$ would imply that $\lb^{mingal}$ is larger than the virial radius of that halo at the current epoch, which means in essence that this halo could not have formed in the first place. Now, our assumption that there is only \textit{one} species of DM bosons of given mass $m$ means that we cannot turn the wheel on $m$ any longer. On the other hand, as we have discussed above, bigger halos than that ultrafaint galactic one have to have a virial radius (much) larger than $\lb^{mingal}$. These halos need to be held up with ``dispersion pressure'', stemming from large-scale dynamics. The envelope of these bigger halos will be constituted of BEC-DM drops (subhalos in CDM nomenclature) of a certain size and mass distribution. Evaluating (\ref{deB})  but accounting for different halo virial velocities, we get a lower bound on the size of drops within any given halo, and that size will differ from $\lb^{mingal}$ (remember the dependence of $\lb$ on environment). For the above choice of $m$, we simply have  
\beq \label{subdeB}
\lb^{subMW} = 1.2~\text{kpc} \left(\f{10 ~\text{km}/s}{v}\right),
\eeq
and around the solar circle and beyond, with $v \approx 200$ km/s, this yields
$\lb^{subMW} \approx 60$ pc for the equilibrium radius of the ground-state BEC-DM drops. The corresponding coherence, or crossing time is $\Tlb^{subMW} \approx 6000$ yr. Of course, the Compton scales are fixed by $m$, which for the above choice amount to $\lambda_0 = 0.04$ pc and $T_0 = 0.132$ yr. \\
At first sight, the above estimates seem counterintuitive, because from CDM we are used to thinking of the common hierarchical structure-formation scenario, where the smallest subhalos form first, merge, and build up ever bigger halos, albeit subhalos can be stripped in the process, as well, and any galaxy today, whether ultrafaint dwarf or Milky Way should have formed by hierarchical build-up\footnote{It is just that the Jeans scale of CDM is much smaller; the smallest CDM subhalos form near the thermal cutoff scale of density perturbations via monolithic collapse, while bigger halos have experienced mergers subsequently.}. This picture is still similar above the minimum Jeans scale of BEC-DM, as simulation results show, but the very mass-radius relationship of ground-state equilibria in (\ref{massradius}), $R \sim 1/M$, is to be blamed for the somewhat counterintuitive picture that virialized DM drops \textit{today} within big hosts like the Milky Way can be smaller than the smallest-size, virialized galactic halo \tx{today}. The inverse dependence of $\lb$ on virial velocity of a given ``halo environment'' is related to the above mass-radius relation, and that relation is obeyed for any ground-state equilibrium; for the above discussion, the fiducial ultrafaint galactic halo was assumed to be such a ground-state equilibrium today, while the Milky Way was \tx{not}. Instead, the Milky Way halo is big and has to have an envelope within which smaller DM drops can ``float around''. By the same token, the halo core of the Milky Way - understood as being close to a ground-state equilibrium (``soliton'') as simulations suggest, has also an equilibrium radius of order $\lb^{subMW} \approx 60$ pc for the fixed boson mass of $m=10^{-21}$ eV/$c^2$. In practice, again, the solitonic core differs from $\lb^{subMW}$ by factors of a few.\\\
The mass of the ground-state drops within the Milky Way can be estimated, using the mass-radius relationship in (\ref{massradius}), and setting $R_{99} = \lb^{subMW}$ for simplicity,
resulting in 
\beq \label{masssubMW}
M^{subMW}(v) \approx 7\times 10^6 ~\MS \left(\f{10^{-21} ~\text{eV}/c^2}{m}\right)\left(\f{10 ~\text{km}/s}{v}\right)^{-1},
\eeq
where we re-introduced the $m$-dependence for an easier comparison later. Inserting $m=10^{-21}$ eV$/c^2$, we have $M^{subMW}(200) \approx 1.4\times 10^8~\MS$ in the solar neighborhood. Notice that this mass is higher by many orders of magnitude, compared to the values  in (\ref{Huimass}) and (\ref{BarOrmass}), once our fiducial choice is picked, and this is a result of observing the nonlinear overdensity of virialized solitonic drops.

We may reconsider the relaxation time of \cite{Hui17} in (\ref{Huirelax}), but now insert (\ref{masssubMW}), instead of their $m_\text{eff}$. Using the same local DM density, we have
\beq \label{relaxnew}
t_\text{relax}(r) \approx 5.7\times 10^8 \text{yr} \left(\f{r}{\rm{kpc}}\right)^4 \left(\f{10^{-21} ~\text{eV}/c^2}{m}\right)^{-1},
\eeq
and the $v$-dependence dropped out, because of the relationship between $\lb, R_{99}$ and $M$.
We can see that the relaxation time for drops is much shorter than the one for wavelets, hence their dynamical impact should be studied in more detail in the future.

We proceed by deriving an admittedly crude upper bound $N^{subMW}(v)$ on the number of such DM drops by assuming that they make up the total mass of the Milky Way today (this is the opposite extreme to the assumption that the halo is made up entirely of wavelets, instead). Taking $M_{MW} = 10^{12}~\MS$, we simply set
$N^{subMW}(v) \approx 10^{12}~\MS/M^{subMW}(v)$.
For our fiducial choice of $m=10^{-21}$ eV$/c^2$ (i.e. demanding that the above ultrafaint galactic halo is a solitonic equilibrium today), and picking $v=200$ km/s, we have $N^{subMW} (200) \approx 10,000$ drops, each of mass $\sim 10^8~\MS$ and radius $\sim 60$ pc.  

It is tempting to compare the upper bound on drops for smaller boson masses $m$; setting again $v=200$ km/s but varying $m$ we have: for $m=10^{-22} ~\text{eV}/c^2$, $N^{subMW} (200) \sim 1000$ drops of mass $\sim 10^9~\MS$ and radius $\sim 600$ pc; for $m=10^{-23} ~\text{eV}/c^2$, we would have only $N^{subMW} (200) \sim 100$ drops of mass $\sim 10^{10}~\MS$ and radius $\sim 6$ kpc. Apart from the implications that would arise, including e.g. limits on disk heating that would follow, such small $m$ would have made impossible the aformentioned ultrafaint galactic halos to exist as solitonic equilibria today, hence small-$m$ models would be ruled out by this simple argument, let alone the accurate bounds described in Sec.1. Higher $m$, respective smaller $\lb^{mingal}$ are allowed, however, because that would just say that the above fiducial ultrafaint dwarf galactic halos are not made up of a pure solitonic equilibrium, but that they also have to have an envelope around a solitonic core, similarly to the Milky Way, but just a very much smaller envelope. In that case, solitonic equilibria could make up smaller-scale halo structures, possibly hosting even smaller galaxies, if they were eventually found.  
In any case, the number of BEC-DM drops rises quickly with increasing $m$, e.g. for $m=10^{-20} ~\text{eV}/c^2$, $N^{subMW} (200) \sim 100,000$ drops of mass $\sim 10^7~\MS$ and radius $\sim 6$ pc.
Before we discuss more implications, let us comment on the roughness of these number estimates. They neglect any nonlinear dynamics like mergers or tidal disruptions.

It has been pointed out that FDM drops are more easily tidally disrupted than CDM subhalos, if the tidal radius of a drop lies within the solitonic core of the primary halo. So, tidal stripping depends sensitively on the position of the drop within its host. 
However, the higher $m$, the smaller the solitonic core. Also, at higher $m$, drops within a halo background can be considered relatively isolated for most of their lifetime, because their size is small compared to their mutual distance. Since ground-state drops have undergone gravitational cooling (as described in Sec.2), we might argue that these drops are little vulnerable against nonlinear ``post-processing''. Of course, there ought to be unbound DM within the Milky Way halo, which arises from the ``cooling'' processes of these small-scale drops. Their expelled, unbound matter will undergo interference akin to the large-scale wave phenomena described above; however, the dynamical importance of that unbound DM becomes diminished, the higher $m$, as mentioned before.

The ever growing importance of drops, with their increasingly smaller size as the boson mass increases, gets them into a regime where they may eventually become MACHO-like  (MACHO: massive, astrophysical, compact halo object).
In the early days of CDM, theoretical studies investigated the dynamical impact of such objects in the Milky Way, whether they be of baryonic origin (e.g. dim stars such as brown dwarfs), or black holes (BHs), or other exotic compact objects which could account for the DM. Later, it has been shown that MACHOs could not be made of baryonic matter, see \cite{Freese2000}. Nevertheless, the issue of compact "dark" halo objects is recurring and observational limits continue to be placed. Given that CDM substructure due to WIMPs, or other DM substructure, is expected to be plentiful, the issue of its impact in the solar neighborhood, or generally in the galactic disk and halo, has not gone away, only because certain MACHOs have come out of favour. And in fact, this impact may be an issue for alternative DM models, once they predict a lot of substructure. For BEC-DM, we will discuss some further potential constraints in the next subsection.

\subsection{Impact onto stellar tracers in the disk and halo and the GAIA era}

It has been long realized that the study of the impact of DM substructure onto stellar tracers, such as stellar clusters, wide binaries, or more recently stellar streams, has the potential to constrain the nature of DM. 
For instance, early work by \cite{BHT85} has shown that the mass of ``dark halo objects'' can be constrained by considerations of wide binary lifetimes, e.g. they found an upper bound of $2~\MS$ for objects in the disk. Also, for sufficiently compact objects, such as MACHOs, the optical depth of microlensing events due to foreground substructure within the Milky Way towards, say, the Magellanic clouds, has been studied over the years.  
The duration of a magnification (microlensing event) is determined by a combination of lens distance, velocity and mass (size does not enter for assumed point-mass lenses). As a result, mass bounds are normally placed on MACHOs, for given assumption on how much they contribute to the halo DM (whether all of it, or parts of it). ``Classic'' microlensing bounds have been placed in \cite{Alcock, Tisserand}, suggesting that allowed MACHOs have to have a mass $\lesssim 10^{-7}~\MS$ or $\gtrsim 30~\MS$.
More recently, \cite{Brandt16} has conducted a study to see how the survival time of compact ultrafaint dwarf galaxies, as well as -again- the survival of the star cluster in Eridanus II, could impact the bounds on MACHOs, finding that any mass in the range $\sim (20-100) ~\MS$ be excluded. Therefore, it is argued that the combined previous constraints from microlensing and wide binaries, along with these findings, rule out MACHOs as the primary DM constituent for any mass upward of 
$\sim 10^{-7}~\MS$.

We argue that a reconsideration of similar bounds such as placed on MACHOs, including the application of more accurate halo models, should be applied to other DM substructure, if the latter is compact enough. This is especially true for BEC-DM solitonic drops, since the inverse mass-radius relationship in (\ref{massradius}) implies that the density of such drops grows rapidly with mass, $\rho \sim M^4$. BEC-DM drops can be more compact than CDM/NFW subhalos, and so may be susceptible to similar constraints than MACHOs. In order to get an intuition, let us check some numbers. If we took the MACHO bounds above at face value, we might demand
\beq \label{highMACHO}
M^{subMW} \lesssim 100~\MS\, ~~\mbox{implying}~~ m \gtrsim 10^{-15}~ \text{eV}/c^2,
\eeq
while
\beq \label{lowMACHO}
M^{subMW} \lesssim 10^{-7}~\MS ~~\mbox{implies}~~ m \gtrsim 10^{-6}~ \text{eV}/c^2. 
\eeq
If a constraint as tight as required by (\ref{lowMACHO}) were to hold up in the future, it would suggest that BEC-DM bosons must have a mass similar to the QCD axion. In fact, the latter is in perfect agreement with such a tight constraint on substructure. While we do not discuss the QCD axion in this paper, its substructure including microlensing has been studied recently e.g. in \cite{Davidson16,2017PhRvL.119b1101F,Schiappacasse_2018,Xiao21}.

In fact, there is a recent body of literature which takes up the question of microlensing 
by extended DM substructure/MACHOs, such as NFW subhalos and boson stars\footnote{Boson stars are the relativistic brethren of solitonic drops; they have higher densities and were originally conceived as potential compact objects to fill the gap between BHs and neutron stars. The limiting mass for solitons made up of bosons, above which they collapse to a BH has been derived in \cite{RB69}, and it is given by $M_{max} \simeq 0.633 m_{pl}^2/m \simeq 8.46\times 10^{10}~\MS \left(\f{10^{-21} ~\text{eV}/c^2}{m}\right)$, where $m_{pl}$ is the Planck mass. Comparing this limiting mass to (\ref{masssubMW}), we can see that there is no danger for BEC-DM drops to collapse to BHs (this statement relies on the assumed absence of self-interaction).}, see \cite{2020PhRvD.101h3013C,2020PhRvD.102h3021C}, as the latter serve as lenses within the Milky Way and M31, magnifying extended stellar sources in M31. While the fraction of point-mass DM structures as lenses can be constrained in a broad range of $\sim (10^{-11} - 10) \MS$, the inferred limits become weaker for increasing lens size. For example, if the size which contains $90 \%$ of the mass of the lens is of order $5$ au, boson stars of mass $\sim (10^{-2} - 10) \MS$ can be excluded to make up a substantial fraction of the DM, while for NFW subhalos of same size, a broader mass range of $(\sim 10^{-4} - 10) \MS$ can be excluded to constitute a substantial fraction. 
\\
On the other hand, dMACHOs (composite objects that are made up of dark-sector elementary particles) have been considered in \cite{2020JCAP...09..044B}, but we will not discuss them here.
\\
Now, what is the size of solitonic drops of mass $M^{subMW}$, and how does it compare to "typical" 
MACHO\footnote{The most extreme MACHOs are BHs: The Schwarzschild radius of a BH with mass $M$ in fiducial units is $R_S = 2GM/c^2 \simeq 3\times 10^5 (M/\MS)~\text{cm}$, e.g. $R_S \approx 0.02$ au for BHs of $10^6~\MS$, and BHs of that mass were required in the early study by \cite{LO85} to explain the Galactic DM and be responsible for the heating of stellar disks.} objects?  Evaluating $\lb$ at the solar circle, for the above approximate lower bounds on the boson masses, reveals that the solitonic equilibrium radius for the case in (\ref{highMACHO}) is $\sim 12$ au, while for the case in (\ref{lowMACHO}) it is $\sim 2$ km, thus asteroid-size. Such small substructure is, again, not unheard-of to the QCD axion. In any case, we can see that drops for $m \ggg 10^{-21}$ eV$/c^2$ can well qualify as point-mass\footnote{However, the point-mass approximation becomes worse for solitons of larger size, not just for obvious reasons, but also for the fact that solitonic drops have no compact support; this is why $R_{99}$ in (\ref{massradius}) was introduced in the first place. Hence, there is no well-defined cutoff between the drops' half-mass radius, tidal radius, or impact parameter for its role as a perturber. These cutoffs must be introduced as appropriate for the problem at hand, e.g. the study of \cite{2020PhRvD.102h3021C} suggests that the precise cutoff does not matter in the microlensing signal. } perturbers for many purposes, even though drops are far from reaching the compactness of "typical" MACHOs. It remains to be seen whether more fluffy solitonic drops, compared to boson stars, can be constrained by microlensing in the future.

In the era of high-precision astrometry and photometry, many previously derived observational bounds should be reconsidered, in general, especially those which attempt to determine the impact of DM substructure (MACHO-like, or other) onto open clusters, wide binaries and stellar streams.
The GAIA mission is currently revolutionizing our census of stellar populations and clusters in the Milky Way, revealing a highly complicated phase space structure within stretches of the Milky Way, devoid of symmetries or isotropy so cherished by theoreticians. New open clusters are being found and previous would-be clusters are being debunked. Many young open clusters are found with large tidal tails $\gtrsim 100$ pc, also for known ones such as the Hyades, see \cite{MA19}, questioning whether we understand their formation and dynamics. A very recent review of the science highlights of the GAIA mission can be found in \cite{GAIA}.

In the spirit of earlier theoretical studies, however, \cite{Hui17} infer some rough estimates on the disruption time of open clusters due to FDM quasiparticles/wavelets, assuming the diffusive regime for extended perturbers. For chosen nominal cluster values of mass $300~\MS$ and size $2$ pc, they find that the disruption time in the solar neighborhood and for $m=10^{-22}$ eV$/c^2$ is of order $2\times 10^{11}$ yr, i.e. too long to be of interest, given cluster ages of $\sim (3-5)\times 10^8$ yr. However, this bound does not apply to BEC-DM/FDM drops. We have only roughly estimated their mass and numbers above, and for our fiducial value of $m=10^{-21}$ eV$/c^2$. 
Comparing the above derived numbers to the roughly 1000 GMCs of mass $\gtrsim 10^5~\MS$ and half-mass radius $r_h \approx 10$ pc, we do well in expecting that, whatever dynamical damage to, say, open clusters can be caused by GMCs, it can be also caused by such drops. It is common lore in \cite{BT2008} that rough estimates of the disruption time of open clusters by GMCc seem to work out to explain the age limit of the former. But the impact of drops onto open clusters would be more prominent, if we believe the rough numbers that we derived.

Now, \cite{Hui17} draw the same conclusion of negligible disruption rates by fiducial FDM wavelets, when it comes to wide binaries (semi-major axis of $a \lesssim 0.1$ pc $\sim 2\times 10^4$ au). Wide binaries are a particularly useful probe, for their wide separation makes them vulnerable to dynamic perturbations. As pointed out in \cite{BT2008}, binaries with $a \gtrsim 0.1$ pc cannot survive easily in the solar neighborhood for a lifetime of $\sim 10$ Gyr, due to the combined effects of high-speed encounters with GMCs and stars. The situation is different in the halo, because the relative velocity dispersion of binaries is much higher than in the disk, and because they spend only a fraction of their Galactic orbit within the disk. However, how to square this ``neat'' state-of-affairs with the supposed DM substructure which is plentiful and small, the higher $m$. Again, it will be interesting to study this problem theoretically and observationally in detail. With respect to the latter, GAIA is also a game changer for the sheer number of binaries it can provide to the community. For instance, \cite{HL20} study around $99,000$ pairs of wide binaries within a distance of $1$ kpc and conclude that those with large separation $>10^4-10^5$ au correspond to the exponential tail of the ``normal'' wide binary distribution, or be chance alignments. A fraction of wide binaries are triples which complicates the analysis. On the other hand, \cite{Tian20} study subsamples of disk, halo and intermediate wide binaries, all within a distance of $4$ kpc, and find that the separation distribution funtion is indistinguishable at small separations, $a < 10^4$ au ($\sim 0.12$ pc), i.e. it follows a single powerlaw, while there occurs a break away from a single powerlaw at higher separations. However, the steepness of the break is different for the different subsamples: the distribution falls off quite steeper for halo binaries than for disk binaries. For one thing, this suggests that the tidal influence by GMCs is not the main contributor to set the outer slope in the separation of wide disk binaries.     
The authors proceed by estimating which MACHO perturbers could be responsible for the break, assuming that the separation distribution was initially a single powerlaw. For solar mass binaries, they find that MACHOs should have a mass $M > 10~\MS$ in order for the disruption time to be smaller than $10$ Gyr.
The authors conclude that this lower limit is in conflict with the upper limits of \cite{Brandt16}, mentioned above which excludes the range $\sim (20-100)~\MS$. Therefore, the assumption of an initial single powerlaw is relaxed with discussions about potentially alternative binary formation scenarios.
Again, if those bounds hold up, they would only leave room for higher-mass bosons (\ref{lowMACHO}). 
However, recent literature questions some of the earlier limits placed on MACHOs, argueing that a large window of intermediate-mass MACHOs with $(100-100,000)~ \MS$ might constitute the DM after all; e.g. with ``old'', new candidates being primordial intermediate-mass BHs, see \cite{CF20}. This mass range would favour drops in models with $10^{-19} \lesssim m/(\text{eV}/c^2) \lesssim 10^{-17}$, though their radial extent with $0.6 \gtrsim r/\text{pc} \gtrsim 6\times 10^{-3}$ exceeds typical MACHO sizes (all quantities are evaluated at the solar circle). 

To reiterate, it will be important in the future to reconsider previous bounds on DM substructure, especially in light of new GAIA data. While it had seemed that fiducial FDM with $m \sim 10^{-22}$ eV$/c^2$, with its grandious exemplification of quantum-coherent behaviour on galactic scales, was originally well-suited to explain the small-scale issues of CDM, new bounds and theoretical studies have increasingly pushed the boson mass to higher values. From the substructure point of view, this may be worrisome because it could imply that there remains a regime of BEC-DM models, without self-interaction, which is prone to produce substructure that can cause too much dynamical chaos in the Milky Way. In order to avoid it, we might have to restrain to a higher regime in $m$ altogether, close to the QCD axion. It might well be that the implied substructure of the QCD axion could stand the hail of upcoming observational bounds raining down on theoretical models. Unfortunately, the wave nature of the QCD axion, while still macroscopic from a laboratory perspective (see Sec.2), is too tiny to make an effect on structure formation and halo dynamics, though there are notable studies to the contrary, see \cite{SY09,Erken12}.

\section{Other novelties}

\subsection{Scalar field oscillations}

The dynamical differences of BEC-DM which behaves as a quantum fluid, compared to CDM, arise from the astronomically large de Broglie scales (\ref{char1}-\ref{char2}).
However, the Compton scales in (\ref{compton}-\ref{char3}) give also rise to effects. While these scales are much smaller than the de Broglie scales, they can still be prominent, especially for BEC-DM with small $m$ (see Sec. 2; $m = 10^{-21}$ eV$/c^2$ has an oscillation time of $ T_0 \approx 50$ days).  This is in contrast to CDM, whose Compton scales will not make an effect on astronomical observations.
The question arises whether these inherent oscillations of BEC-DM as a scalar field can be potentially observed\footnote{All papers discussed in this section share the assumption that plane-wave approximations for the wave function can be used. This is justified, so long as the spacetime curvature does not vary significantly over a region of the size of the spatial spread of the (quantum) particle.}, and this question has been answered positively in the work by \cite{KR14}, as follows. 
The underlying scalar field oscillates with a frequency that depends on fundamental parameters, such as mass $m$, but also on other potential contributions such as boson self-interaction. However, the contribution due to the rest-mass (\ref{compton}-\ref{char3}) dominates in any case.
These scalar-field oscillations are passed on to other observables, notably the (hydrodynamical) pressure which, like the energy density, can be derived from a general-relativistic stress-energy tensor of BEC-DM.  The amplitude of the oscillating pressure averages to zero (for time scales much longer than $T_0$ !), and this is the reason why BEC-DM behaves like CDM on cosmological scales. However, those oscillations in pressure induce oscillations in gravitational potentials, and as such these resulting oscillations act similarly to passing-by gravitational waves, which also cause perturbations of gravitational fields. Therefore, photons which pass through such oscillating gravitational fields will likewise suffer shifts in arrival times. As a result, this effect can be searched for
 using radio telescope networks for pulsar timing arrays such as NANOGrav \cite{Nanograv}, which can indirectly infer the presence of oscillating scalar fields through a measurement of coherent shifts in the pulse arrival times of a sample of precisely spinning millisecond pulsars. \cite{KR14} find that the effect on pulsar timing is comparable to the corresponding effect of a monochromatic gravitational wave with characteristic strain $h_c \approx 2\times 10^{-15} \left(\f{10^{-23}~\text{eV}/c^2}{m}\right)^2$ and frequency $f = 5\times 10^{-9}~\text{Hz} \left(\f{m}{10^{-23}~\text{eV}/c^2}\right)$. In terms of the magnitude of the signal, it is comparable to the stochastic gravitational wave background produced by massive BH binary systems.
 
 However, no signal from oscillating BEC-DM has yet been detected. The Parkes Pulsar Timing Array (PPTA) has placed upper limits on the local DM density for ultralight BEC-DM/FDM with $m \lesssim 10^{-23}$ eV$/c^2$ in \cite{PPTA}. Back then, data was used of timing observations for 26 pulsars over a decade. Probing BEC-DM of higher $m \gtrsim 10^{-22}$ eV/$c^2$ is challenging and will require the monitoring of hundreds of pulsars to high precision for more than a decade. If pulsars happen to be so close to each other that they come to lie within the same de Broglie length ``patch'' of a DM wavelet or drop, phase-coherent oscillations arise as shown in \cite{Schive17}, which can be also searched for.  Of course, the higher $m$, the smaller $\lb$, and the likelihood of finding pulsars within this distance becomes very small. In any case, finding and monitoring millisecond pulsars toward the Galactic center has a great potential to place bounds on BEC-DM models.   

 However, the analysis in \cite{KR14} applies to real scalar fields, appropriate for axion-like particles. A similar analysis, but for complex scalar fields, reveals that the expected signal is very much smaller\footnote{In the past, such differences with respect to oscillation amplitudes have been also noticed for complex boson stars versus real oscillons.} than that for the real case \cite{bachelor}, placing the observational limits far beyond any future measurement capabilities.

\subsection{Coherence and quantum measurement in Astronomy}

In this last subsection, we touch upon some issues that have implications for our understanding of quantum mechanics and quantum gravity.

We have mentioned quantum interference phenomena of BEC-DM, which is supposed to be in a coherent state on cosmic scales, so let us introduce some notions first.
Coherence describes all the properties of the correlation between physical quantities of a single wave, or between several waves or wave packets. The degree of coherence is mathematically described via correlation functions. A perfectly coherent state has a density matrix that is a projection onto the pure coherent state, and it is equivalent to a wave function, while a mixed state is described by a classical probablity function for the pure states that make up the mixture. Now, the ``wave function'' of the BEC-DM condensate (i.e. the single scalar field that describes an $N$-particle system) is supposed to have this property of coherence.
It has been shown that there is a correspondence to quantum entanglement, i.e. quantum coherence is equivalent to quantum entanglement in the sense that coherence can be described as an entanglement measure, see \cite{quant}, and conversely each entanglement measure corresponds to a coherence measure.
With this in mind, cosmic BEC-DM should be quantum-entangled, but that means cosmic BEC-DM is not ``separable'': any parts can occupy any of their possible states and in order to explain the behaviour as a whole, one has to consider all possible states. It is only once a quantum measurement is performed on a part that its state is determined, but so will instantly be the ``rest''. This is common lore in a laboratory setting, but lifting this up to the cosmic scale, it would imply that a local measurement, e.g. within the Milky Way halo, would instantly determine the ``outcome'' or state of BEC-DM in the rest of the Universe!   

What would qualify as a (quantum) measurement ? The indirect dynamical inference of BEC-DM onto baryonic tracers?  Do baryonic tracers somehow act as an environment for BEC-DM, which makes the latter decohere\footnote{A related problem has been considered in \cite{Hertzberg1,Hertzberg2}, where two DM particles form a Schr\"odinger-cat-like system, interacting with an incoming baryonic particle. However, the case of a macroscopic BEC-DM system is not covered by their analysis, though a rough estimate suggests that the decoherence rate for boson stars is very rapid, for all particle masses of interest here. This reinforces earlier work by \cite{GHC15}, which concludes that such boson stars behave ``classical enough'' to justify an effective field description.}?
If not, can cosmic BEC-DM be regarded as an ``isolated system'', for it pervades the whole Universe anyway?
Would the direct detection of a boson (axion-like or other) constitute that particular measurement that would determine the cosmic state of coherent, entangled BEC-DM?
Is quantum measurement questioned by this problem?

This topic has been addressed by the beautiful paper of \cite{Helfer}. It takes FDM with boson mass of order $10^{-22}$ eV$/c^2$ at face value, which has a Compton length of order $\lesssim 1$ pc. While an effective classical description is adequate above the Compton scales, around and below it such a description would break down (this has been known). However, in addition, issues of quantum measurement theory come into play, which have not previously been pointed out. At and below the Compton scale, measurements imply the excitation of relativistic modes, even if the state is initially nonrelativistic, and such states brought about by the measurement have a uniform distribution up to the wave number set by the measuring scale. As a result, a substantial fraction of modes will be relativistic upon measurement, and would be able to escape the Milky Way. FDM would be unstable against quantum measurements, as Helfer \cite{Helfer} puts it and ``the stability of the Galaxy is hostage against such observations''.  But even if one argues that this known issue is not an issue, there is a further point to consider and this is the main focus of \cite{Helfer}.  

Quantum measurements generally violate conservation laws. In a laboratory setting, this problem can be put under the rug, because any failure of conservation in the observed system can be thought to be absorbed by the much bigger measurement apparatus, i.e. the failures can be made small, if not eliminated within conventional quantum theory. But in the case of BEC-DM/FDM with large Compton length, the energies involved in the measuring devices can be small, compared to the system - which is e.g. the local DM density. Thus, a DM candidate such as BEC-DM challenges our (non-)understanding of quantum measurement theory. Since DM interacts gravitationally, the observables to probe quantum measurement derive from a stress-energy tensor, i.e. they are related to geometry and, in essence, entail implications for quantum gravity, too.  Now, \cite{Helfer} considers quantum fields described by a special-relativistic Klein-Gordon equation (upon ``second quantization''). A renormalized stress-energy operator is calculated from which energy density and pressure of FDM can be derived. These quantites are all averaged over a length scale a few times the Compton scale. The interference of individual modes and their energy is determined. By defining a quantum-coherent state of ``many possibilities'', the measurement of, say, the averaged energy density will project that state onto an eigenspace, since the measurement ``picks'' one possibility. It can be shown that for both energy density and pressure, the quantum measurement will excite relativistic modes (even if initially nonrelativistic), i.e. particles will depopulate out of their states, become gravitationally unbound, and escape the Milky Way (or any galaxy or cluster, if the measurement would be made there). In order to show that such a measurement could be performed in principle (if not by humans at this time), \cite{Helfer} moves on to present a particular scenario. It is conceptually similar to the gravitational-wave detection via pulsar timing arrays, but cannot rely on pulsars because the measurement has to be performed within a Compton scale of $\lesssim 1$ pc. Instead of two world lines, one of the emitter (pulsar) and one of the receiver (observer on Earth), the timing scheme employs a single world line of an emitter, which sends out s-wave pulses according to its proper time $\tau_e$, that are reflected from a sphere of mirrors at their proper time $\tau_r$, along world lines which are stationary (at 0th order), at a particular distance from the emitter, and the signals are subsequently received on the original world line (that of the emitter). If the distance between emitter and receiver is $D/2$, the required time resolution to measure the quantum corrections to the difference $\tau_e - \tau_r$ is around $10^{-7}$ sec, if $D \approx 1$ pc. If $D$ is smaller, the requirement on resolution increases.

This proposal highlights that there are issues with quantum measurement that have become exposed with BEC-DM. An obvious resolution could be that quantum measurement theory, or our understanding of conventional quantum mechanics is incomplete (which certainly is). Of course, another resolution would be to say there ought to be no BEC-DM, which would seem a ``bad excuse'' argument, in our opinion.
\cite{Helfer} adds a discussion, one of which concerns alternative approaches to quantum theory, e.g. the de Broglie-Bohm formulation, whose formalism was also sketched in Sec.2. However, the latter still lacks a generally accepted relativistic extension. It is certainly true that by using the de Broglie-Bohm formulation with its quantum hydrodynamical equations, astrophysicists have put certain aspects under the rug. On the other hand, if this approach allows us to infer limits on DM - in the form of BEC-DM, axion-like particles, etc. - by studying its impact on real-world baryons, there is hope that this endevaour may shed more light onto fundamental questions than originally expected, i.e. not only on the DM conundrum, but also on issues in quantum mechanics and quantum gravity.

\section{Conclusion}
 
We have reviewed certain novel aspects of dark matter in the form of a Bose-Einstein condensate (BEC-DM), and how its new quantum phenomena may be studied and constrained within our own Milky Way.
While the literature has continuously placed more and tighter limits on the mass of the boson of BEC-DM, there are still many probes left which should be studied in the future, on the theory side and also on the observational side, given the ongoing GAIA revolution. 
Quantum phenomena will be increasingly harder to infer, the higher the boson mass, and ever smaller-scale probes within the Milky Way will have to be used. In this paper, we have disregarded any microphysical extensions, notably boson self-interactions, and so similar studies than those conducted for BEC-DM without self-interaction will be required, in order to see whether any BEC-DM model will reveal itself one day as the alternative solution to the CDM paradigm.



\section*{Conflict of Interest Statement}

The author declares that the research was conducted in the absence of any commercial or financial relationships that could be construed as a potential conflict of interest.



\section*{Funding}
The author is supported by the Austrian Science Fund FWF through an Elise Richter fellowship, grant nr. V656-N28.

\section*{Acknowledgments}
The author thanks Stefan Meingast for helpful discussions.

\bibliographystyle{frontiersinHLTH&FPHY} 
\bibliography{DM_frontiers_REV}{}


\end{document}